\documentclass[aps,pre,preprint,superscriptaddress,twocolumn,showpacs,10pt]{revtex4-1}
\usepackage{graphicx}

\begin{document}


\title{Dynamics of short one-dimensional nonlinear thermostated atomic chains}


\author{A. N. Artemov}
\email[]{artemov@fti.dn.ua}
\affiliation{Donetsk Institute of Physics and Engineering, Donetsk
 83114, Ukraine}


\date{\today}

\begin{abstract}
The dynamics of short 1D nonlinear Hamiltonian chains is analyzed
numerically at different temperatures (energy per particle). The
boundary temperature $T_b$ separating the regular (quasiperiodic)
and the stochastic (chaotic) chain motion is found. The dynamical
properties of short 1D nonlinear chains interacting with
thermostats are studied. It is shown that, in spite of the
fluctuations, the dynamics of such systems can be stochastic as
well as regular. The boundary temperature of these systems is
close to that of the Hamiltonian one.
\end{abstract}


\maketitle

\section{Introduction}
{\label{I}}

Fermi et. al \cite{Fermi} (FPU) were first who began to study
numerically statistical properties of nonlinear 1D atomic chains.
The reason for the failure of their attempt to get the
redistribution of the energy of one mode between others due to
nonlinearity now is clear \cite{Ford}.

Kolmogorov \cite{Kolm}, Arnol'd \cite{Arnold} and Moser
\cite{Moser} (KAM theorem) have proofed that in the case of small
nonintegrable addition term to integrable Hamiltonian the system
under some conditions can hold quasiperiodic motion on an
invariant torus. Izrailev and Chirikov \cite{Chirik1} supposed
that there is the border of stochasticity for nonlinear chains.
They used the resonance overlap criterion and obtained analytical
condition of the onset of chaos. Further numerical calculations of
nonlinear Hamiltonian chains \cite{Chirik2,Diana} confirmed these
results.

Now the study of the chains which are in the contact with
thermostat is of most interest. Dynamical properties of such
systems are more complicated. They are not Hamiltonian ones and
describe the open systems which can exchange energy with
environment. The systems under uniform temperatures are of
marginal interest. But the systems with nonuniform temperature are
much more interesting because they provide the thermal transport
from higher temperature to lower. Such systems are investigated
intensively.

The different dynamical properties of different 1D chains result
in various heat transport laws. The heat transfer in macroscopic
solids obeys the Fourier heat conduction law
\begin{equation}\label{e1.1}
    J=-k\nabla T,
\end{equation}
where $J$ is the heat flux, $k$ is the thermal conductivity, which
doesn't depend on a sample size, and $\nabla T$ is the temperature
gradient.

Now the validity of the Fourier law is justified by means of
numerical calculations in the ding-a-ling \cite{Casati} and the
Frenkel and Kontorova \cite{Hu} models. The integrable systems,
such as harmonic \cite{Rieder} and Toda \cite{Hu2} chains, don't
form the temperature gradient because there is no scattering of
the thermal excitations which transport the heat. The thermal
conductivity of these systems is divergent. The FPU chains
\cite{Kabur, Lepri, Fill, Das}, the diatomic Toda lattice
\cite{Hatano}, the Heisenberg spin chain \cite{Dhar} take
intermediate states. The thermal conductivity of the models
diverges as $k\sim N^\alpha$ at the number of particles
$N\rightarrow \infty$ with $0<\alpha <1$.

In this paper we do not consider the thermal conductivity of
chains with the large numbers of particles. We set sparer task
which is to examine the dynamical behavior of short chains of the
FPU type, which are in the contact with thermostates, and to
compare them with that of similar Hamiltonian chains under
suitable conditions.

\section{Hamiltonian chains}
\label{II}

The dynamical properties of the systems with great number of the
degrees of freedom are hard to investigate numerically. To avoid
the problem we consider the atomic chains with small numbers of
particles and assume that their characteristic properties can be
extended to the chains with greater dimension.

As a model we used the 1D chain of $N$ classical particles with
mass $m$ positioned in the points $x_i$ and interacting with each
other via a nonharmonic potential $U(x_{i+1}-x_i)$. In the
equilibrium state the particles are located at equal distances $a$
from each other. Dynamics of the chain obeys the classical
equations of motion
\begin{eqnarray}
  m\frac{d^2x_i}{d t^2} &=& -\frac{1}{2}\frac{d}{d
  x_i}\left[U(x_i-x_{i-1})+U(x_{i+1}-x_i)\right],
  \label{e2.1}
\end{eqnarray}
where $i=1..N$. Indexes $i=0$ and $N+1$ correspond to the
motionless points with coordinates $x_0=0$ and $x_{N+1}=(N+1)a$
which are the boundary conditions for Eq.(\ref{e2.1}). Potential
used in the model is the sum of the harmonic and the quartic terms
\begin{equation}\label{e2.2}
    U(x)=\frac{k}{2}x^2+\frac{\beta}{4}x^4,
\end{equation}
where $k$ is the elasticity factor and $\beta$ is the nonlinearity
parameter.

Hereinafter the model parameters used are $m=k=a=1$ and $\beta=2$.
The system of the ordinary differential equations Eq.(\ref{e2.1})
was solved numerically by means of the modification \cite{Hockney}
of the Verlet algorithm \cite{Verlet}.

The randomly chosen particle velocities at the equilibrium
particle positions were used as the initial conditions. The
kinetic energy per particle, which we call "temperature", is used
as driving parameter which controls the type of dynamics.

\begin{figure}[h]
  \includegraphics[width=8.5cm]{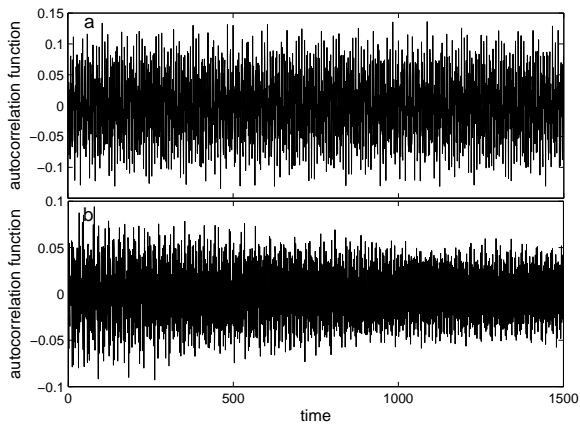}\\
  \caption{Autocorrelation functions (a) of the  quasiperiodic motion of
  4-particle chain at $T=0.131$ and  (b) of the chaotic one at $T=0.133$. }
  \label{f1}
\end{figure}

\begin{figure}[h]
  \includegraphics[width=8.5cm]{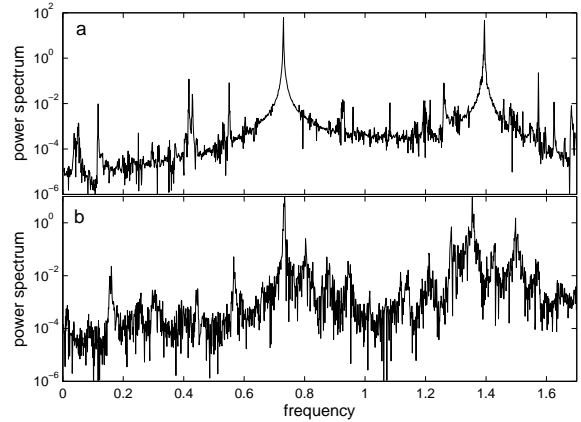}\\
  \caption{Power spectrum (a) of the  quasiperiodic motion of
  4-particle chain at $T=0.131$ and  (b) of the chaotic one at $T=0.133$. }
  \label{f2}
\end{figure}

Now it is known that a nonlinear Hamiltonian particle chain with
$N>1$ can demonstrate a quasiperiodic or a chaotic type of motion.
To identify the character of the chain motion we analyzed the time
series, which are the sets of velocities of the second particle in
a chain taken with the time interval $Dt=1$. Such an approach is
widely used to investigate the properties of chaotic systems
\cite{Schuster}. The power spectrums and the autocorrelation
functions of the solutions were the criteria of the chain
dynamics. The examples of these functions corresponding to the
4-particle chain which is in a regular and a chaotic dynamical
states in the close vicinity of the boundary temperature are shown
in Figs.\ref{f1} and \ref{f2}.

\begin{figure}[h]
  \includegraphics[width=8.5cm]{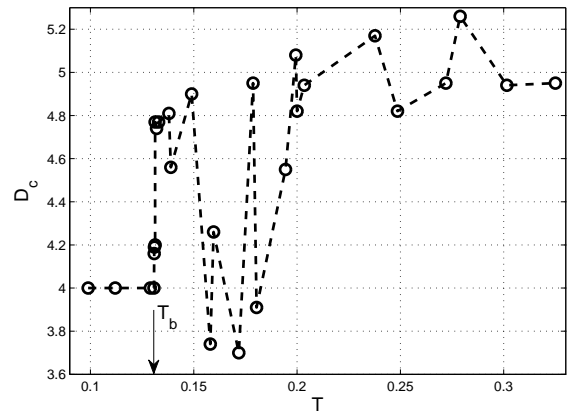}\\
  \caption{\label{f3}Correlation dimension $D_c$ versus temperature $T$
  for the 4-particle chain. The arrow points the boundary temperature
  between the regular and the chaotic motions}
\end{figure}

An additional information was obtained by means of the calculation
of the correlation dimension $D_c$ which evaluates the lower limit
of the Hausdorff dimension of the subset of the phase space
occupied with the solutions. In the case of a quasiperiodic motion
$D_c$ gives the dimension of the corresponding invariant torus. In
all cases examined $D_c=N$ for the regular dynamics. In the chaos
regime the correlation dimension seemingly strongly depends on
initial conditions and is a random value. Temperature dependence
of the correlation dimension of the 4-particle chain solutions is
plotted in Fig.\ref{f3}.

If these characteristics did not make it possible for
identification of the type of motion unambiguously, the time
behavior of phase trajectories with close initial conditions was
analyzed.

\begin{figure}
 \includegraphics[width=8.5cm]{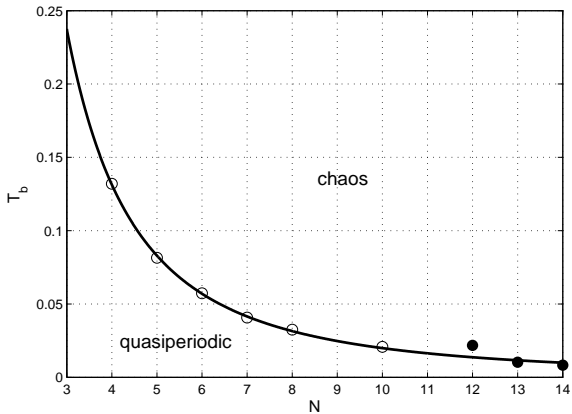}%
 \caption{Temperature $T_b$ separating two types of dynamics as a function of the chain
  length $N$. Open circles show  the boundary temperatures of the Hamiltonian chains.
  Line is the fitting curve which follows the dependence
  $T_b(N)=2.28\cdot N^{-2.06}$. The close circles show
  $T_b$ of the thermostated chains.\label{f4}}
 \end{figure}
We examined the behavior of the chains with $N=2-8,10$ at
different initial conditions in the regions of a regular and a
chaotic motion to find the boundary temperature $T_b$ between
them. Results of the calculations are shown in Fig.\ref{f4}. The
numerically obtained boundary temperatures are marked by open
circles. These points separate the regions of a regular and a
chaotic dynamics of the chains. The dependence obtained was fitted
by the function $T_c=2.28 N^{-2.06}$ shown by the solid line. The
numerical parameters were found by means of the least square
method. This dependence is in qualitative accordance with the
Chirikov criterion \cite{Chirik1} which predicts that critical
parameter $\epsilon_s\sim N^{-2}$ for large $N$ and high modes.

There are two short chains ($N=2,3$) which fall out from the
dependence discussed. Two-particle chain under reasonable
restriction on temperature (particle shift $\triangle x\lesssim
a$) demonstrates a regular dynamics only. In the case $N=3$
$T_b\approx 0.965$ is considerably larger than that in
Fig.\ref{f4}.

As it follows from the obtained dependence, the boundary
temperature $T_b$ decreases rapidly when the chain length $N$
grows. Moreover, the total boundary energy of the chains
$E_N=NT_b$ decreases as $N^{-1}$.

\section{Thermostated chains}
\label{III}

In this section we consider the 1D atomic chains interacting with
a thermostat, i.e. with a large system which is under constant
temperature and isn't influenced by the chain. This problem is
very different from that considered in the previous section.

The motion of the Hamiltonian chains is stipulated unambiguously
by the initial conditions and the chain total energy is an
integral of motion. In this case the system dynamical properties
can be obtained by means of analyzing the phase trajectories or
the time series.

An open system exchanges energy with a thermostat. In this case
the system total energy is fluctuating. Other difference is that
the dynamics and fluctuations of the system are driven by the same
cause, that is the thermostat temperature. In this case the
contributions of the dynamics and the fluctuations in the motion
of the system are the same and we can't separate them.

In this section we consider the chains in which the first and the
last particles are in the contact with the thermostats with
temperatures $T_1$ and $T_N$. The thermostat temperatures are the
parameters of the problem. We used both the Langeven and the
N\'{o}se-Hoover \cite{Hoover} thermostats. Both of them give the
qualitatively same results. Therefore we show the results obtained
with the N\'{o}se-Hoover thermostat only.

\begin{figure}[h]
  \includegraphics[width=8.5cm]{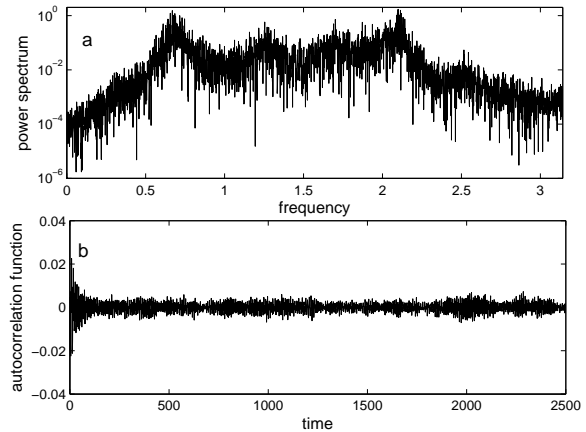}\\
  \caption{Power spectrum (a) and autocorrelation function (b) of
  the 4-particle chain at the thermostat temperature $T=0.05$.}
  \label{f5}
\end{figure}

First we considered the chains at the uniform temperature
$T=T_1=T_N$. To understand the chain behavior we analyzed the time
series and plotted the autocorrelation functions and the power
spectrums of the chains with different $N$ and the thermostat
temperatures $T$. The results for $N=4$ and $T=0.05$ are shown in
Fig.\ref{f5}. As it is seen in Fig.\ref{f4} this temperature
corresponds to the quasiperiodic motion of the Hamiltonian chain.
However, the plots Fig.\ref{f5} are characteristic rather to noise
with short correlation time than to a dynamical motion. Such a
picture is typical for the chains with other $N$ and at the
temperatures regions corresponding to both a quasiperiodic and a
chaotic motion.

Other approaches used in Section \ref{II} also don't allow to
identify a type of motion. The correlation dimensions are more
than the total phase space one $D_c>2N$ at any temperatures what
doesn't correspond to either dynamical motion. The behavior of two
phase trajectories with close initial conditions at any
temperatures is typical to a developed chaos.

\begin{figure}[h]
  \includegraphics[width=8.5cm]{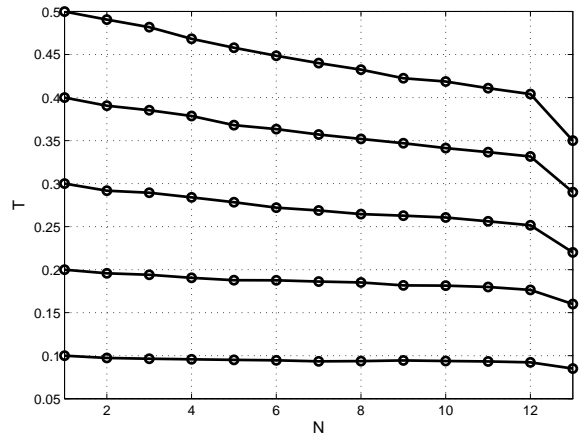}\\
  \caption{Temperature profiles of the 13-particle chain.}
  \label{f6}
\end{figure}

But the study of the chains with a nonuniform temperature
distribution was more informative. We calculated the temperature
distribution along the chains at different temperatures on left
and right ends and plotted the temperature gradients as functions
of temperature for different chains. The main difficulty in such
an approach is that the temperature gradient at a fixed number of
particles strongly depends on both temperatures. We tried to find
the maximum temperature gradient which can be formed in each chain
at given left temperature $T_1$. The temperature profiles for the
chain with $N=13$ at left temperatures in the interval
$T_1=0.1-0.5$ are plotted in Fig.\ref{f6}. To find the profile we
specified the thermostat temperatures on the chain ends and waited
during time interval $t=10^{4}$ to thermalize the chain. Then we
get the local temperature as a particle kinetic energy averaged
over the time interval $t=10^{5}$.

\begin{figure}[h]
  \includegraphics[width=8.5cm]{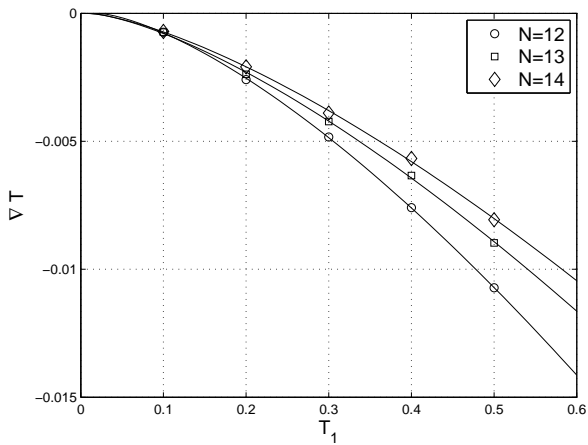}\\
 \caption{\label{f7}Temperature gradient as function of temperature $T_1$
 for the chains of different length.}
\end{figure}

The temperature gradients as the functions of the temperature of
the first particle $T_1$ are shown in Fig.\ref{f7} for the chains
with $N=12,13,14$. They were found by the least square method
using $N-1$ points of the chains with the exception of the last
one. The markers show the points calculated. The solid lines fit
the dependencies obtained using the expression
\begin{equation}\label{e3.1}
    \nabla T=A(T_1-T_b)^b.
\end{equation}
The fitting parameters are shown in the Table. The parameter $T_b$
is the temperature at which the temperature gradient becomes zero.
At lower temperatures $T<T_b$ the temperature gradient is equal to
zero.

\begin{center}
\begin{tabular}{|c|c|c|c|} \hline
$N$ & $T_b$ & $A$ & $b$  \\  \hline 12 & 0.0213 & -0.0315 & 1.46 \\
\hline  13 & 0.0103 & -0.0248 & 1.43 \\ \hline 14 & 0.0083 &
-0.0221 & 1.42
\\ \hline
\end{tabular}
\end{center}

\begin{figure}[ht]
  \includegraphics[width=8.5cm]{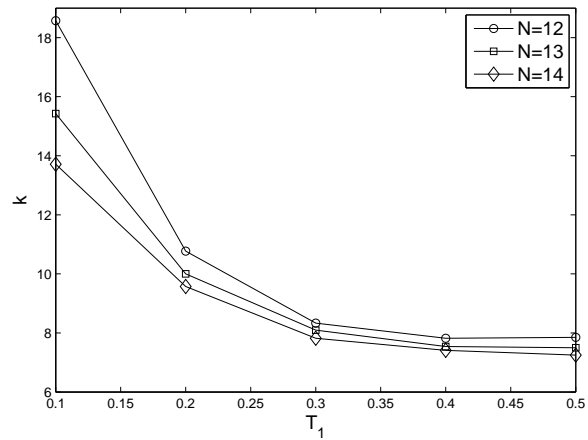}\\
  \caption{\label{f8}Temperature dependence of the thermal conductivity for the chains of different length.}
\end{figure}

The heat flux is finite at any temperature $T_1$ if $\triangle
T=T_1-T_N$ is finite. Therefore the thermal conductivity diverges
as soon as $T_1\rightarrow T_b+0$ and it is divergent at $T<T_b$.
This means that the interaction of the thermal excitations
transferring the heat, which is responsible for the conductivity,
disappears at this temperature. The thermal conductivity is
plotted as the function of the temperature $T_1$ in Fig. \ref{f8}.

The temperatures $T_b$ of the thermostated chains are shown in
Fig. \ref{f4} by the close circles. They lie in the close vicinity
of the positions corresponding to the boundary temperatures $T_b$
of the Hamiltonian chains with the same number of particles. These
temperatures separate the regions of stochastic and the regular
motion of the chains under thermal fluctuations. So, we can see
analogy between the  dynamics of the Hamiltonian and the
thermostated chains.

\section{Conclusion}

We considered the dynamics of short Hamiltonian chains and chains
interacting with the thermostats. It is shown that both types of
the chains demonstrate similar dynamical behavior. In particular,
the Hamiltonian chains depending on temperature can be in two
dynamical states, regular (quasiperiodic) and stochastic
(chaotic), which are separated by the boundary temperature $T_b$
(Fig.\ref{f4}). The results were obtained by means of analyzing of
time series.

This approach doesn't work in the case of the thermostated chains
because of strong thermal fluctuations. In this case we used the
indirect method to ascertain dynamics of the chains. We analyzed
the temperature dependence of the slope of the temperature
profiles of the chains and found the temperature $T_b$ at which
the slope vanishes. At the temperatures $T\leq T_b$ the thermal
conductivity is divergent as it takes place in integrable systems.
So, we can suppose that in this temperature interval the chain, in
spite of the fluctuations, demonstrates the regular (fluctuated
quasiperiodic) motion. Accordingly, at $T>T_b$ the chain dynamics
is stochastic (fluctuated chaotic).

The method used to analyze the thermostated chains allows us to
evaluate the boundary temperatures only. We can specify a regular
or a stochastic chain motion but can't obtain its more detailed
characteristics, which are responsible for the thermal
conductivity behavior. Apparently, for this purpose investigation
of the Hamiltonian chains is more promising.


\end{document}